 \documentclass[apj]{emulateapj}
\usepackage{epsfig}

\def\nlam{32 }
\def\nhot{12 }
\def\nltten{14 }
\def\ntwosig{19 }
\def\Ronemode{1.34}

\def\kozfmax{0.64}
\def\kozlowlim{0.40}
\def\kozhilim{0.87}
\def\scatfmax{0.55}
\def\scatlowlim{0.34}
\def\scathilim{0.76}
\def\twomodefinterpretation{Both models indicate a significant fraction of aligned systems, with the PSTF model indicating that nearly half of systems might be intrinsically aligned.  So the next question to ask is in this two-channel picture of planet migration, which model do the data favor?}

\def\twomodeR{-1.16}
\def\twomodeconf{94\%}
\def\twomodemodel{PSTF }
\def\twomodemodelselectinterpretation{Thus, while the current sample of $\lambda$ measurements appears at first to be evidence for Kozai migration as predicted by FT07, the model preference changes in favor of the N08 scattering model if allowance is made for the existence of some systems forming hot Jupiters via some mechanism that preserves alignment.  And since what current RM measurements can tell us about planet migration mechanisms changes significantly depending on whether we assume one or two migration channels, we next explore whether current data allows us to distinguish between those scenarios.    
}

\def\onevtwoconf{99\%}
\def\onevtworesults{We calculate $\mathcal R=-3.00$, which strongly favors the PSTF + aligned model over the one-mode KCTF model, with a confidence of $\onevtwoconf$ (Figure \ref{fig:onevtwoall}). }

\def\hotonemodeconf{82\%}
\def\hotstarresults{Of our two one-mode migration models, PSTF is favored over KCTF, with $\hotonemodeconf$ confidence ($\mathcal R=-0.73$; Figure \ref{fig:confhotone}).  We also explore the two-mode migration hypothesis, allowing for hot stars to also have an intrinsically aligned population.  The posterior distributions for $f$ under the two models conditioned on the hot star data are illustrated in Figure \ref{fig:fhot}, showing that both models favor almost complete misalignment.  In fact, one-mode vs.~two-mode model selection for this subset of the data gives $\mathcal R=-0.33$, favoring the single-mode hypothesis.  For this reason we do not pursue any further the idea that there exists an intrinsically aligned population among hot stars, as the current data do not merit the additional model complication. 

}

\def\howmanyresults{We learn from this that in order to have a good chance of obtaining a hot star data set that confidently selects either of our single-mode migration models, we will likely need a total data set of about 80-100 hot star RM observations (Figure \ref{fig:howmany}). More optimistically, there is $>50\%$ chance of confident model selection with only a factor of $\sim$3 more observations (40 total) if the PSTF model is correct.  Given the pace at which this field is growing, this may be reasonably expected to happen within the next few years.  On the other hand, if the KCTF model (or some other model) is a better description of reality, then it will likely take more observations to determine this, given the preference of the current data for the PSTF model.

}

\begin{document}

\title{Discerning Exoplanet Migration Models Using Spin-Orbit Measurements}
\author{Timothy D.~Morton and John Asher Johnson}

\begin{abstract}
We investigate the current sample of exoplanet spin-orbit measurements to determine whether a dominant planet migration channel can be identified, and at what confidence.  We use the predictions of Kozai migration plus tidal friction \citep{fab07} and planet-planet scattering \citep{nagasawa08} as our misalignment models, and we allow for a fraction of intrinsically aligned systems, explainable by disk migration.  Bayesian model comparison demonstrates that the current sample of \nlam spin-orbit measurements strongly favors a two-mode migration scenario combining planet-planet scattering and disk migration over a single-mode Kozai migration scenario.  Our analysis indicates that between $34\%$ and $76\%$ of close-in planets ($95\%$ confidence) migrated via planet-planet scattering.  Separately analyzing the subsample of \nhot stars with $T_{\rm eff} > 6250$ K---which \citet{winn10} predict to be the only type of stars to maintain their primordial misalignments---we find that the data favor a single-mode scattering model over Kozai with $85\%$ confidence.  We also assess the number of additional hot star spin-orbit measurements that will likely be necessary to provide a more confident model selection, finding that an additional 20-30 measurements has a $>50\%$ chance of resulting in a $95\%$-confident model selection, if the current model selection is correct.  While we test only the predictions of particular Kozai and scattering migration models in this work, our methods may be used to test the predictions of any other spin-orbit misaligning mechanism.

\end{abstract}

\section{Introduction}

Exoplanets that transit their host stars provide opportunity to study
distant planetary systems in great detail.  Most immediately,
photometry during transit measures a planet's radius and density, but follow-up studies can provide
much more. Secondary eclipse photometry, 
photometric phase curve measurements, and transmission spectroscopy, for
example, can reveal a planet's temperature, albedo, atmospheric
composition, and even weather patterns.  While these tools investigate
the physical characteristics of the planets themselves, transits
 also provide a valuable opportunity to measure details of
planets' 
orbital characteristics using the Rossiter-McLaughlin (RM) effect. 

The RM effect is an anomalous Doppler signal due to the shadow of a
transiting planet crossing the face of a rotating star, and is
measured by obtaining radial velocity (RV) measurements of the host
star during transit.  As the approaching limb of the stellar surface
is occulted, the total integrated radial velocity is redshifted, and
as the receding limb is occulted, the integrated velocity is
blue-shifted.  Modeling the RM effect yields a measure of the angle
between the orbital axis of the planet and the projected rotation axis
of the host star, typically referred to as $\lambda$.   While this
angle is not itself physically meaningful, it constrains
$\psi$, the true angle between the planet's orbit and the star's
rotation. In addition to being a fundamental system parameter akin to
semimajor axis or eccentricity,  $\psi$ is a potential window into
learning about planetary orbital migration, as different migration
scenarios predict different distributions of $\psi$.   

The first several RM measurements that were made all indicated small
values of $\lambda$ \citep{winn05,winn06b,wolf07,narita07}, and thus were consistent with small values of
$\psi$, which was not unexpected since the orbits of all the planets
in the Solar System are aligned to within $7^\circ$ of the solar spin
axis.  The first misaligned system, XO-3, was discovered by \citet{hebrard08}, and confirmed by \citet{winn09a}. Since then, many additional misaligned systems have been discovered \citep{winn09b,johnson09,triaud10}.  This diversity of measured $\lambda$s suggests that planetary
migration is more complicated than simple disk migration, which 
predicts planet orbits well aligned with stellar spins (unless the disk itself is misaligned),
and even hints at multiple migration channels. 

As the number of exoplanet systems with measured values of $\lambda$ increases, so does the desire and ability to draw conclusions based on the data.  For example, \citet{fw09} (FW09) suggest that there might be two distinct populations of close-in planets---intrinsically aligned and intrinsically misaligned---with a 95\% probability of $>64\%$ of planets belonging to the aligned population.  This remarkable result was based on the first 10 RM measurements, which included a single misaligned system. 

There now exists a sample of  \nlam published spin-orbit angles, which provides a valuable opportunity to readdress and extend this previous study, particularly in light of the recent proposition by \citet{triaud10} that current data suggest that all hot Jupiters migrated via the Kozai mechanism \citep{kozai,wu07,fab07}.   The central goal of this paper is to determine whether the current sample of spin-orbit measurements is sufficient to begin to discern between the predictions of different exoplanet migration theories such as the Kozai mechanism and planet-planet scattering \citep{nagasawa08}, and if not, then how many more RM observations will be needed to draw more meaningful conclusions about the intrinsic $\psi$ distribution of transiting exoplanets.  


We describe our models in \S\ref{models} and our analysis in
\S\ref{analysis} and \S\ref{twomode}.   In  \S\ref{hotstars} we repeat our analysis
restricted to hot 
stars (following the suggestion of \citet{winn10}).
In \S\ref{howmany}, we look 
to the future and ask how many RM observations will be needed to draw
more confident statistical conclusions.  We conclude our discussion in
\S\ref{discussion}.    

\section{The Models}
\label{models}

We test two hypotheses against each other in this paper.  The first is that close-in exoplanets migrated to their present-day orbital locations through a combination of the effects of Kozai cycles and tidal friction, as described by \citet{fab07} (FT07).  Kozai cycles are oscillations in eccentricity and inclination of a close binary system caused by the presence of distant third companion in an inclined orbit.  If a giant planet (assumed to have formed beyond the ice line in its protoplanetary disk) undergoes Kozai cycles that cause it to pass within a few stellar radii of its host star, then tidal friction can quickly circularize its orbit, freezing in a potentially large orbital inclination ($\psi$) to the newly-migrated hot Jupiter.  FT07 uses 1000 simulations of such systems,using Jupiter-mass planets with initial orbital separations of 5 AU and outer 1 $M_\odot$ companions on a 500 AU orbits highly inclined with respect to the planet's initial orbit.  The final inclination of the resulting close-in planets in these simulations is their prediction of the hot Jupiter $\psi$ distribution resulting from this migration mechanism (FT07, Figure 10).  In the rest of the paper, we refer to this model as ``KCTF.''

The second hypothesis is that planet-planet scattering is the dominant mechanism for forming close-in planets, as modeled by \citet{nagasawa08} (N08).  In their simulations, they study the evolution of systems of three Jupiter-mass planets with initial orbital separations of 5.00, 7.25 and 9.50 AU and initial inclinations of 0.5$^\circ$, 1.0$^\circ$ and 1.5$^\circ$.  In addition to planet-planet scattering, they also include the effects of the Kozai mechanism and tidal friction, and produce a final population of close-in planets with a large range of orbital inclinations (N08, Figure 11(c)).   We refer to this model as ``PSTF.'' 

There have been many different simulations similar in nature to these that we test (e.g.~\citet{wu07,chatterjee08,ford08,juric08}), but these are the two that produce the broadest distribution of $\psi$ values, and thus seem most likely to be able to explain the observed retrograde systems.  In addition, \citet{triaud10} also compared the observed data with these two models, concluding that the KCTF model of FT07 describes the current data better than PSTF of N08.

\begin{figure}[!t]
	\epsscale{1.3}
	\plotone{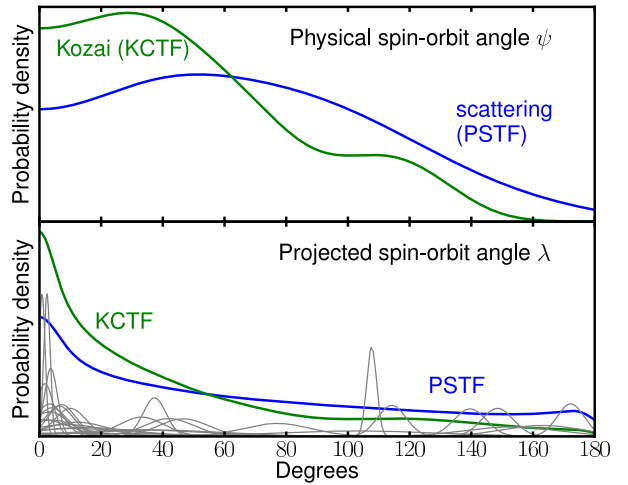}
	\caption{Probability distributions for the true ($\psi$; top
          panel) and projected ($\lambda$; bottom panel) spin-orbit
          angles that our two misalignment mechanisms produce.  The
          KCTF $\psi$ distribution is taken from the simulations of
          final inclinations of planets migrating through Kozai cycles
          + tidal damping from \citet{fab07}, and the planet-planet scattering (PSTF) prediction is taken from the simulations of \citet{nagasawa08}.  We use
          Monte Carlo simulations to project each $\psi$ distribution
          to create the $\lambda$ distributions.  
          Note that non-zero
          but small $\lambda$ values are preferred by
          the KCTF model, while large values ($\lambda > 60^\circ$) are preferred
          by the PSTF model.  The bottom panel shows the distribution of the \nlam measured $\lambda$ values.\label{fig:psilambda}} 
\end{figure}

We emphasize that while the true spin-orbit angle $\psi$ is the physically meaningful angle, only the projected version of this angle $\lambda$ is measurable through the RM effect, because the inclination of the stellar rotation axis is unknown.  If the star is rotating completely edge-on ($I_\star = 90^\circ$), then $\lambda = \psi$, but in general the star's rotation axis may be tilted along the line of sight, resulting in $\lambda \neq \psi$ (see FW09, Fig.~3).   In other words, observation of a large value of $\lambda$ is firmly indicative of a large $\psi$, but observation of a small $\lambda$ does not rule out the possibility of large $\psi$.  In rare cases, this unknown stellar inclination can be constrained by combining photometric determination of a rotation period with an estimate of the stellar radius and rotational line broadening \citep{winnholman07}.  Additionally, the likelihood for any particular transiting planet to be misaligned may be estimated before even any RM measurement occurs, by comparing the observed line broadening to theoretical rotation predictions \citep{schlaufman10}.  In general however, since the inclinations of individual stellar rotations are unknown, statistical methods must be employed to draw conclusions about $\psi$ from an observed population of $\lambda$.   

One strategy to do this, employed by \citet{triaud10}, is to statistically deproject each $\lambda$ measurement to form a posterior probability distribution function (PDF) for $\psi$ for each system, assuming an isotropic distribution for the inclination of stellar spins, and sum them to form an overall $\psi$ PDF.  However, this method has the disadvantage that isotropy of stellar spins is actually a strong assumption, since the observed distribution of spin inclinations will depend on the true $\psi$ distribution, which we are trying to determine.  This is analogous to how the orbit inclinations of RV-detected planets may not always be assumed to be isotropically distributed, as discussed in \citet{ho10}.  

Our analysis relies instead on comparing the observed $\lambda$ data directly to theoretical predictions of the models, in $\lambda$-space.  This requires that we determine a probability density function (PDF) for $\lambda$ corresponding to each of the $\psi$ distributions we wish to test.  Since both of our $\psi$ predictions are the results of complicated simulations, we use a Monte Carlo procedure to perform this transformation.

We fix an observer-oriented spherical coordinate system and simulate a large number of systems as follows. First, we populate $10^6$ stars with transiting planets with orbital inclination vectors $\vec O$ assigned according to the distribution of known transiting planet inclinations.   Then the stellar spins $\vec S$ are assigned relative to the planet population, according to the predicted $\psi$ distributions of KCTF and PSTF.  This is accomplished by selecting a $\psi$ from the distribution implied by the simulations of FT07 or N08, treating this $\psi$ as a polar angle relative to $\vec O$, and assigning $\vec S$ to have an azimuthal angle around $\vec O$, chosen uniformly on ($0,2\pi$).  Then $\vec S$ is transformed back into the observer-oriented coordinate system, in which $\lambda$ is simply the azimuthal angle of $\vec S$ about the line of sight.  The probability distribution for $\lambda$ is then determined from the distribution of these resulting $\lambda$ values.  Figure~\ref{fig:psilambda} illustrates the original $\psi$ and derived $\lambda$ distributions for both misalignment models, as well as the current $\lambda$ data.

In addition to comparing these two misalignment models, we also consider the possibility that there might be multiple migration mechanisms, inspired by the conclusions of FW09, who found the data available at the time (10 systems) favored a model with planets drawn from two distinct distributions: one perfectly aligned ($\psi=0$) and one isotropically misaligned.  After introducing our methods in \S\ref{analysis}, we explore in \S\ref{twomode} what we can learn if we assume only a fraction $f$ of systems are misaligned according to one of the above mechanisms, with the remaining $1-f$ fraction being perfectly aligned.

\section{Data Analysis}
\label{analysis}

The goal of our analysis is to select the model that describes the data best, and to determine the confidence with which we can make that selection.  We do this first assuming that all planets are misaligned, and then again allowing for an aligned population.  The following sub-sections outline the details of these steps.  The data we use are the 28 RM measurements compiled in Table 1 of \citet{winn10} (with 5 of these systems updated; Winn, in prep), plus HAT-P-14 (185$^\circ \pm 4.5$), (Winn, in prep), HAT-P-4 (-15$^\circ \pm 16$) (Winn, in prep), and XO-4  \citep{narita10}.  A summary of the results of all our analysis is in Table 1.


\subsection{Which misalignment mechanism is preferred?}
\label{whichmodel}

We use the Bayes factor (e.g. ?) for our model selection statistic, which in the simple case of comparing two models with no free parameters reduces to a likelihood ratio, as follows:
\begin{equation}
\label{Rdef}
\mathcal R = \log_{10}\left(\frac{\mathcal L_{\rm Koz}(\{\lambda\})}{\mathcal L_{\rm scat}(\{\lambda\})}\right),
\end{equation}
where $\mathcal L_{\rm Koz}(\{\lambda\})$ and $\mathcal L_{\rm scat}(\{\lambda\})$ are likelihoods of the observed data $\{\lambda\}$ under the two different misalignment models.  $\mathcal R > 0$ favors the KCTF model, and $\mathcal R < 0$ favors the PSTF model.  The likelihoods are calculated as follows:
\begin{equation}
\label{likelihooddef}
\mathcal L_{\mathcal M} (\{\lambda\}) = \displaystyle \prod_{i=1}^N P_{\mathcal M}(\lambda_i),
\end{equation}
where
\begin{equation}
\label{poflambda}
P_{\mathcal M}(\lambda_i)  = \displaystyle \int_0^{180} p_i (\lambda) p_{\mathcal M,\lambda}(\lambda) d \lambda,
\end{equation}
with $p_i(\lambda)$ being the probability distribution of the $i$th $\lambda$ measurement (which we take to be a Gaussian centered at the published value with width as the published error bar), and $p_{\mathcal M,\lambda}(\lambda)$ being the $\lambda$ PDF for the model in question ($\mathcal M = \{{\rm Koz,scat}\}$).

Using the current set of \nlam $\lambda$ measurements, we calculate $\mathcal R = \Ronemode$, which favors KCTF over the PSTF.  The following section explains how we quantify the strength of this model selection.

\subsection{Confidence Assessment}
\label{confidence}

Within the dichotomous paradigm of comparing two misalignment models we can determine the confidence in our model selection by answering the following question: ``Given a measured value of $\mathcal R=\Ronemode$, which favors the KCTF model, what is the probability that the KCTF model is actually correct?''   Or, more generally, given any measured value of $\mathcal R$, how can we quantify the confidence in the implied model selection?   

To address this question, we perform the analysis described above on many simulated data sets.  Starting with $10^5$ randomly-generated spin-orbit systems for each misalignment model, constructed according to the procedure described in \S\ref{models}, we randomly select \nlam $\lambda$ values from each underlying model. Each simulated $\lambda$ measurement is the exact value of $\lambda$ drawn from the simulations, perturbed by a measurement error $\sigma_\lambda$, assigned using Eq.~16 from \citet{gaudi07}: 


\begin{equation}
\sigma_\lambda = \frac{\sigma}{v\sin I_\star
  \sqrt{N}}\frac{1-\gamma^2}{\gamma^2}\left[\frac{(1-b^2)\cos^2
    \lambda + 3b^2\cos^2\lambda}{b^2(1-b^2)}\right]^{1/2}, 
\end{equation}

\noindent where $\sigma$ is the single-point RV measurement uncertainty, $v\sin I_\star$ is the projected rotational velocity of the star, $N$ is the number of RV points in transit, $\gamma$ is the planet-star radius ratio, and $b$ is the transit impact parameter.  For these simulated data sets, we randomly assign $v \sin I_\star$, $\gamma$, and $b$ by drawing randomly with replacement from the current set of all transiting planet systems\footnote{as listed at www.exoplanets.org}.  We take $\sigma = 5~{\rm m/s}$ and $N=20$ for each simulated measurement.  

 
 \begin{figure}[!t]
 	\epsscale{1.2}
 	\plotone{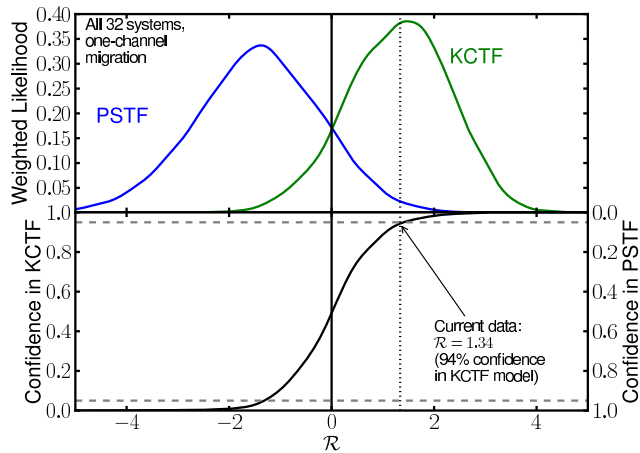}
	\caption{The relationship between our model
         selection statistic $\mathcal R$, or the logarithm of the likelihood ratio, and a model selection confidence.  We
          randomly draw 5000 sets of \nlam $\lambda$ values from each
          model and measure $\mathcal
          R$ for each of these data sets.  The top panel show the
          distribution of $R$ values attained from these data
          simulations.  The confidence in the model selection at
          any particular $\mathcal R$ (bottom panel) is determined by
          the relative heights of the two distributions at that
          $\mathcal R$.  The current data strongly favor the KCTF model. \label{fig:confallone}}
\end{figure}

We repeat this data simulation procedure 5000 times and calculate $\mathcal R$ (Eqn.~\ref{Rdef}) for each data set, giving us an understanding of the expected distribution of $\mathcal R$ if either of these models do describe the actual underlying $\psi$.  Using these simulations to construct PDFs for $\mathcal R$ under each model ($p_{\mathcal R,{\rm Koz}}(\mathcal R)$ and $p_{\mathcal R,{\rm scat}}(\mathcal R)$) we can then ask what the probability is of either model being true, given a measurement of $\mathcal R$.  Applying Bayes' theorem with a uniform prior on which model should be correct, we may write:
 
 \begin{equation}
 \label{pofmodelgeneral}
{\rm Pr}(\mathcal M | \mathcal R) = \frac{{\rm Pr}(\mathcal R | \mathcal M)}{\displaystyle \sum_{\mathcal M} {\rm Pr}(\mathcal R | \mathcal M)} = \frac{p_{\mathcal R,\mathcal M}(\mathcal R)}{\displaystyle \sum_{\mathcal M} p_{\mathcal R,\mathcal M}(\mathcal R)},
\end{equation}
where again $\mathcal M = \{{\rm Koz,scat}\}$.  For our specific case, this becomes:
\begin{equation}
\label{pofmodelspecific}
\label{problam}
{\rm Pr}({\rm Koz} | \mathcal R=\Ronemode) = \frac{p_{\mathcal R,{\rm Koz}}(\Ronemode)}{p_{\mathcal R,{\rm Koz}}(\Ronemode) + p_{\mathcal R,{\rm scat}}(\Ronemode)}  = 0.96,
\end{equation}
giving $96\%$ confidence in the KCTF model.  Figure \ref{fig:confallone} illustrates this confidence-assessment procedure.  

This result appears to support the conclusion of \citet{triaud10}, who claim that the current RM data points to the FT07 KCTF model as explaining the formation of hot Jupiters better than the N08 model.  However, as there are reasons to be skeptical that the Kozai mechanism could plausibly be responsible for the formation of all hot Jupiters, both theoretical \citep{wu07} and empirical \citep{schlaufman10}, we take our analysis one step further and consider what we may conclude if we allow for two distinct populations of systems: aligned and misaligned.

\section{Multiple migration channels?}
\label{twomode}

Of the \nlam $\lambda$ measurements to date, \nltten have $|\lambda| \le10$ and \ntwosig are within $2\sigma$ of $\lambda=0$.  Given that KCTF predicts many more systems with small $\lambda$ than does the PSTF model, it is thus not surprising that it is preferred over PSTF in the above analysis.  But what if only a fraction of planets migrate via a misaligning mechanism while the rest migrate through a process such as disk migration \citep{lin96} that preserves spin-orbit alignment?  Analyzing the first 10 $\lambda$ measurements (that included only a single significantly misaligned system), FW09 found such a two-population model to be their best selection.  Especially given the difficulties of explaining all hot Jupiter migration using the Kozai mechanism alone \citep{wu07}, and that \citet{schlaufman10} found that fewer systems seem to misaligned than would be predicted based on this prediction, it seems prudent to investigate how allowing for an aligned population affects conclusions about the intrinsic $\psi$ distribution.  

Accordingly, we add a parameter to each of our models: $f$, the fraction of systems that are misaligned, with $\psi=0$ (and thus $\lambda=0$).  This requires us to modify our model selection procedure (\S\ref{twomodeselect}), but it also enables us to ask an additional question: what do the models imply about $f$ (\S\ref{fraction})? 


\subsection{What fraction of systems are misaligned?}
\label{fraction}

We may use Bayes' theorem to calculate the probability distribution for the misaligned fraction $f$ for each of our two models, conditioned on the \nlam observed $\lambda$ values.  For this particular case, we may write Bayes' theorem as follows:

\begin{equation}
p_{f,\mathcal M}(f | \{ \lambda \}) = \frac{\mathcal L_{\mathcal M}(\{ \lambda\} | f) p(f)}{\displaystyle \int_0^1 \mathcal L_\mathcal M(\{ \lambda\} | f) p(f) df}
\label{eqn:bayes}
\end{equation}
where as before $\{\lambda\}$ is the set of observed $\lambda$ values and $\mathcal M$ represents a particular model (either `Koz' or `scat'), $p_{f,\mathcal M}(f | \{ \lambda \})$ is the posterior probability distribution for $f$ under model $\mathcal M$, $\mathcal L_{\mathcal M}(\{ \lambda\} | f)$ is the likelihood of the data given a particular $f$ under $\mathcal M$, and $p(f)$ is the prior probability distribution for $f$, which we take to be flat between 0 and 1.  The denominator is the normalization factor, also known as the marginal likelihood.   The posterior probability distribution for $f$ allows us to infer conclusions about $f$ for a particular model given the current data.  The likelihood function is calculated the same way as Eq.~\ref{likelihooddef}, except that now the probability distribution for $\lambda$ is dependent on $f$:
\begin{equation}
p_{\lambda,\mathcal M}(\lambda | f) = f\times p_{\mathcal M}(\lambda) + (1-f)\delta(\lambda),
\label{eqn:kozlam}
\end{equation}

\noindent where $p_{\mathcal M}(\lambda)$ are the $\lambda$ distributions we calculated in \S\ref{models} (bottom panel of Figure \ref{fig:psilambda}), and $\delta(\lambda)$ is the Dirac delta function.

\begin{figure}[!lt]
	\epsscale{1.2}
	\plotone{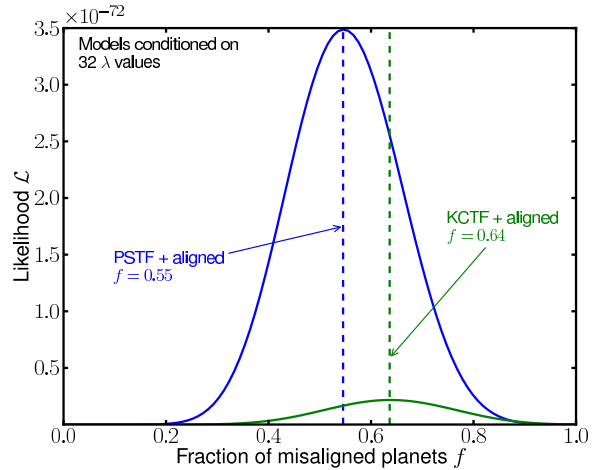}
	\caption{The posterior distribution of $f$ for each of our two-migration-channel
          models, conditioned on \nlam observed $\lambda$ values.  We
          show that if the true misaligned population were misaligned
          according to the KCTF distribution, then we expect that
          about $64\%$ of close-in planets are misaligned.  Similarly,
          if the misaligning mechanism were PSTF, then the most
          likely value of $f$ is close to $55\%$.  Strictly speaking,
          this figure shows the likelihood of the current data as a
          function of $f$ for each model---if these curves were
          normalized they would be proper probability distributions,
          given our flat prior for $f$.  However, plotting them in
          un-normalized form is illustrative, since we use the ratio of the areas under
          these curves (the marginal likelihood) as our two-mode model selection criterion. \label{fig:fall}}
\end{figure}

Fig.~\ref{fig:fall} illustrates $\mathcal L_\mathcal M(\{\lambda\} | f)$ as a function of $f$ for each model.  We measure the most likely values and their  symmetric 95$\%$ confidence ranges for $f$ by normalizing the likelihoods and computing the cumulative distrubution functions. If the KCTF + aligned  model is correct, $f$ lies between $\kozlowlim$ and $\kozhilim$ with 95$\%$ confidence, with the most likely value being $\kozfmax$. Similarly, if the PSTF + aligned model is correct, $f$ lies between $\scatlowlim$ and $\scathilim$ with 95$\%$ confidence, with the most likely value being $\scatfmax$.  \twomodefinterpretation

\subsection{Two-mode model selection and confidence assessment}
\label{twomodeselect}

\begin{figure}[!t]
	\epsscale{1.2}
	\plotone{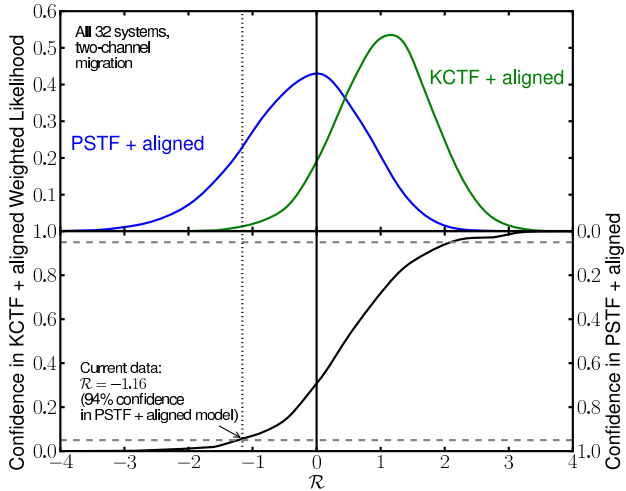}
	\caption{The relationship between our model selection statistic and model selection confidence, for comparing our two-mode migration models, where only a fraction $f$ of planets are misaligned and the rest are in perfectly-aligned systems.  $\mathcal R$ in this case is the logarithm of the ratio of the marginal likelihoods of the two models, and the $\mathcal R$ distributions are generated by measuring the $\mathcal R$ values for data simulations at different $f$.  The PSTF + aligned model is preferred over the KCTF + aligned model. See \S\ref{twomodeselect} for details.
	\label{fig:confalltwo} }
\end{figure}

Since the models we are now comparing each have an unknown parameter $f$, we redefine our model comparison statistic to be the ratio of the marginal likelihoods (the denominator of Eqn.~\ref{eqn:bayes}, or the area under the curves in Figure \ref{fig:fall}) of the two models:
\begin{equation}
\label{Rdeftwomode}
\mathcal R = \log_{10} \left(\frac{\displaystyle \int_0^1 \mathcal L_{\rm Koz}(\{\lambda\} | f) p(f) df}{\displaystyle \int_0^1 \mathcal L_{\rm scat} (\{\lambda\} | f) p(f) df}\right),
\end{equation}
where again $\mathcal R>0$ favors KCTF + aligned and $\mathcal R<0$ favors PSTF + aligned.  This time, we calculate $\mathcal R = \twomodeR$, which favors PSTF over KCTF.  

Confidence assessment requires data simulations as before, except now we simulate data sets for each model on a grid of $f$ values, to determine the behavior of $\mathcal R$ as a function both of misalignment model and $f$.  So for each of 20 equally-spaced values of $f$ between 0 and 1, we randomly draw 1000 sets of 30 systems, as described in \S\ref{confidence}, simulating the aligned fraction of planets by giving each simulated $\lambda$ a probability $(1-f)$ to be re-assigned to $\lambda=0$ before being perturbed by the measurement error.  

These simulated $\mathcal R$ values give us an $\mathcal R$ PDF for each $f$ for each model---essentially empirical two-dimensional likelihood functions: $\mathcal L_{\mathcal R,{\rm Koz}}(\mathcal R,f)$ and $\mathcal L_{\mathcal   R,{\rm scat}}(\mathcal R,f)$, where $\mathcal R$ is continuous but $f$ is only sampled at 20 points between 0 and 1.  Since we already calculated the posterior probability distribution of $f$ given the current data for each of our models above, we may marginalize these likelihood functions over $f$ to calculate a properly weighted one-dimensional PDF for $\mathcal R$: 
\begin{equation}
p_{\mathcal R,\mathcal M} (\mathcal R) = \displaystyle\sum_{i=1}^{20} \mathcal L_{\mathcal R,\mathcal M}(\mathcal R,f_i) {\rm Pr}(f_i | \{\lambda\},\mathcal M).
\end{equation}

The term ${\rm Pr}(f_i |\{\lambda\},\mathcal M)$ is the probability obtained by integrating  $p_{f_i,\mathcal M}(f | \{\lambda\})$ (Eqn.~\ref{eqn:bayes}; Figure~\ref{fig:fall})  between $f_i  - \Delta f/2$ and $f_i + \Delta f/2$, where $\Delta f$ is the spacing between successive $f$ values in our simulations.   Using these likelihood functions, we may calculate the probability of either of our models being true, analogously to Eqns \ref{pofmodelgeneral} and \ref{pofmodelspecific}.  We find that our measurement of $\mathcal R=\twomodeR$ gives a confidence of $\twomodeconf$ for the \twomodemodel model, illustrated in Figure \ref{fig:confalltwo}.

\begin{figure}[!t]
	\epsscale{1.2}
	\plotone{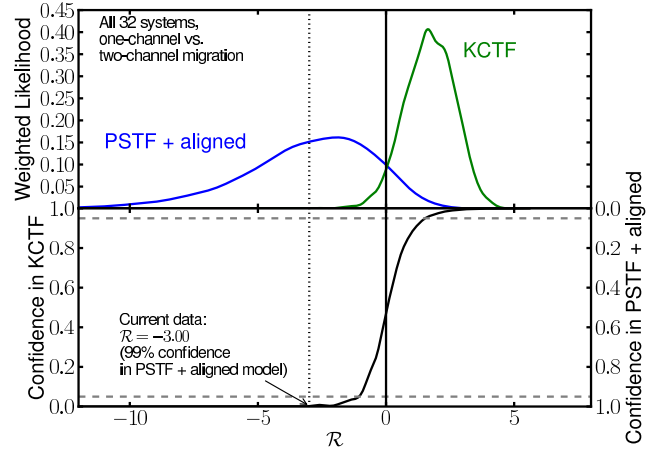}
	\caption{The relationship between our model selection statistic and model selection confidence, comparing the preferred single-mode migration scenario (KCTF) to the preferred two-mode model (PSTF + aligned).   PSTF + aligned is strongly preferred over single-mode KCTF.
	\label{fig:onevtwoall} }
\end{figure}

\twomodemodelselectinterpretation

\subsection{One channel or two channels?}
\label{oneortwo}

We approach this question the same way we have hitherto approached the other model selection questions: define a model selection statistic $\mathcal R$, calculate $\mathcal R$ for the current data, and determine confidence based on data simulations.  Our model selection statistic in this case becomes:
\begin{equation}
\label{Rtwoonedef}
\mathcal R = \log_{10}\left(\frac{\mathcal L_{\rm one}}{\mathcal L_{\rm two}}\right),
\end{equation}
where $\mathcal L_{\rm one}$ is one-mode likelihood as used in Eq.~\ref{Rdef}, and $\mathcal L_{\rm two}$ is a marginalized likelihood as used in Eq.~\ref{Rdeftwomode}.  Since KCTF is preferred for one-mode migration and PSTF is the preferred two-mode model, we take $\mathcal L_{\rm one}$ to be the likelihood of the data under the KCTF model and $\mathcal L_{\rm two}$ to be the marginal likelihood of the data under the PSTF+aligned model.  
\onevtworesults

\section{Hot Stars}
\label{hotstars}

\begin{figure}[!t]
	\epsscale{1.2}
	\plotone{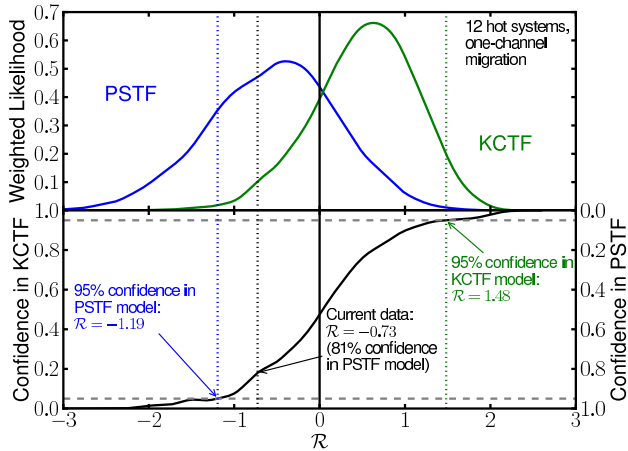}
	\caption{The relationship between our model selection statistic and model selection confidence, but using only the \nhot hot stars ($T_{\rm eff}> 6250$) as our data.  One-channel migration favors the PSTF mechanism for this subset of the data, but not conclusively.  The $\mathcal R$ thresholds required to reach $95\%$ confidence in either model for this sample size are marked.\label{fig:confhotone} }
\end{figure}

We have shown that the KCTF prediction for the distribution of $\psi$ for migrated planets does not adequately explain the observed distribution of $\lambda$, and that current data favor a combination of well aligned systems and systems misaligned via planet-planet scattering.  However, it may be that the current population we observe is not representative of the initial misalignment distribution.  For example, \citet{matsumura10} conclude that tidal effects may be important in damping out initial misalignment in some systems.  \citet{winn10} note that most of the misaligned planets that have been observed to date are around stars hotter than 6250 K.  Based on this empirical finding, they suggest that perhaps all HJs migrate through some misaligning mechanism, and that cooler stars with deep convective zones have their envelopes tidally torqued into alignment by the planet on a relatively  quick timescale, thereby erasing the evidence of the initial misalignment.  If this were indeed the case, then an important key to understanding HJ migration would be spin-orbit measurements of planets  transiting hot stars, since these systems would presumably have maintained their primordial misalignment.   

Following this line of inquiry, we repeat the analyses of \S3 and \S4 restricted to only the current sample of \nhot hot stars ($T_{\rm eff} > 6250$ K).  \hotstarresults

Thus, if the distribution of hot star $\lambda$ values represents the primordial alignment distribution for all stars, the current \nhot $\lambda$ observations hint that the PSTF prediction of N08 describes planet migration better than the KCTF model of FT07. However, there are not yet enough data for this model selection to be conclusive.  In the next section, we discuss how many more RM measurements will likely be needed in order to make hot star model selections  confident.

\begin{figure}[!t]
	\epsscale{1.2}
	\plotone{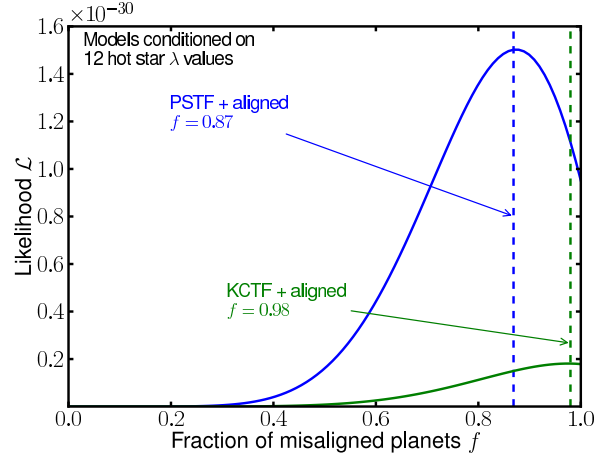}
	\caption{The posterior distribution of $f$ for each of our two-migration-channel
          models, conditioned on the \nhot observed hot star ($T_{\rm eff}> 6250$) $\lambda$ values.  As \citet{winn10} point out, hot stars appear to be misaligned more often than the overall stellar population.  Because $f=1$ has a high probability under both models, we do not further pursue two-mode migration model comparison for this subsample. \label{fig:fhot} }
\end{figure}

\section{How many hot star RM observations are needed?}
\label{howmany}

With the full sample of $\nlam$ RM observations to date, we are able to confidently state that a combination of well aligned systems and systems misaligned via planet-planet scattering (the PSTF model) explains the current data better than migration via the Kozai mechanism and tidal friction alone (KCTF).  Only considering the sample of $\nhot$ hot stars, our present conclusions are weaker.  Inspired by the work of \citet{swiftbeaumont}, who calculated the sample size of protostellar cores necessary to reliably distinguish a power law from a lognormal distribution, we wish to quantify how many additional hot star $\lambda$ measurements will likely be needed in order to measure $\mathcal R$ values indicative of confident ($>95\%$) model selection between the KCTF and PSTF models.  

This requires determining two things: first what values of $\mathcal R$ correspond to $95\%$ confident model selections for a given sample size $N$, and secondly how likely we are to actually measure such a confident value once we have $N$ observations in hand, given the current set of $\nhot$ as our starting point.  

Both questions can be addressed by data simulations.  The $\mathcal R$ thresholds that represent $95\%$ confidence for a given $N$ can be determined by simulating many data sets of sample size $N$ and using Eq.~\ref{pofmodelgeneral} to calculate model selection confidence as a function of $\mathcal R$.  This has already been illustrated for $N=\nlam$ in Figures \ref{fig:confallone}, \ref{fig:confalltwo}, and \ref{fig:onevtwoall} and for $N=\nhot$ in Figure \ref{fig:confhotone}.  We repeat this procedure for different sample sizes, defining $\mathcal R$ thresholds for our model selection for a series of $N$ values up to $N=100$. 

\begin{figure}[!t]
	\epsscale{1.2}
	\plotone{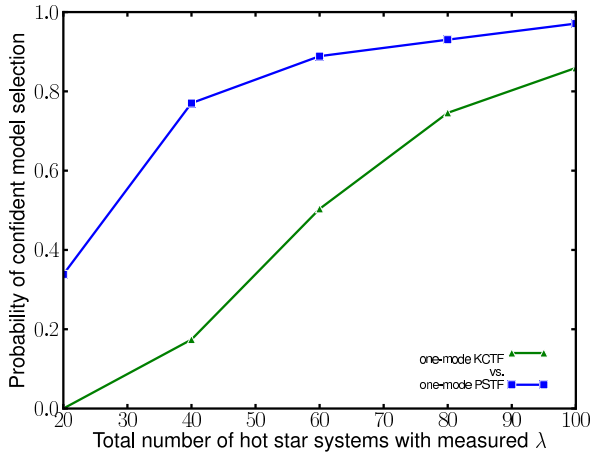}
	\caption{  The probability with which a given-sized future sample of hot-star ($T_{\rm eff} > 6250$ K)          $\lambda$ values (in addition to the \nhot that currently exist) will result in a confident selection ($>95\%$) of one
          of our misalignment models over another.  We use data simulations of
          various sample sizes (described in \S\ref{howmany}) to define the
          confidence thresholds and to determine how likely those
          thresholds are to be reached for each particular sample size.  For example, with 28 more measurements (40 total), we have a $75\%$ chance of selecting the PSTF model at $>95\%$ confidence. 
          \label{fig:howmany}	}
\end{figure}

We also use data simulations to determine the probability of a future experiment resulting in a confident $\mathcal R$ measurement.  To do this we again simulate many data sets of the same sizes as above and measure the $\mathcal R$ distributions of the experiments, but this time the first $\nhot$ measurements in each data set are fixed to be the current hot star measurements.  Then we use the $\mathcal R$ distributions to predict how often threshold values are reached for each model.  This probability is plotted against sample size in Figure \ref{fig:howmany}.  

\howmanyresults

\section{Discussion and Conclusions}
\label{discussion}

\begin{deluxetable*}{ccccccc}
	\tablecolumns{10}
	\tablecaption{Results summary}
	\tablehead{ \colhead{Data} & \colhead{Model 1} & \colhead{Model 2} & \colhead{$f$\tablenotemark{a} (model 1)} & \colhead{$f$\tablenotemark{a} (model 2)} & \colhead{$\mathcal R$\tablenotemark{b}} & \colhead{Confidence\tablenotemark{c}}}
	\startdata
All \nlam systems& \textbf{KCTF\tablenotemark{d}} & PSTF\tablenotemark{e} &  ---  & ---  & \Ronemode & $96\%$ \nl
All \nlam systems & KCTF + aligned & \textbf{PSTF + aligned} & $0.64\pm 0.24$ & $0.55\pm 0.21 $ & \twomodeR & $93\%$ \nl
All \nlam systems & KCTF & \textbf{ PSTF + aligned} & --- & $0.55\pm 0.21 $ & -3.00 & $99\%$ \\ \\ \hline \\
\nhot hot systems\tablenotemark{f}& KCTF & \textbf{PSTF} &  ---  & ---  & -0.73 & $81\%$ \\
	\enddata
	\tablenotetext{a}{maximum-likelihood value, with symmetric $95\%$ confidence range; only applicable to two-mode models}
	\tablenotetext{b}{$\mathcal R > 0$ favors model 1; $\mathcal R < 0$ favors model 2}
	\tablenotetext{c}{Degree of belief that the model selection is correct; based on Monte Carlo simulations (\S3.2)}
	\tablenotetext{d}{Kozai cycles + tidal friction \citep{fab07}}
	\tablenotetext{e}{Planet-planet scattering + Kozai cycles and tidal friction \citep{nagasawa08}}
	\tablenotetext{f}{host star $T_{\rm eff} > 6250$ K}
\end{deluxetable*}

The study of exoplanet spin-orbit angles is advancing rapidly, with \nlam measured projected spin-orbit angles and more to come as ground- and space-based surveys continue to detect transiting planets.  We have analyzed the current sample and quantify what may be inferred about the distribution of true spin-orbit angles $\psi$.  In particular, we ask whether the current data are sufficient to test the predicted distribution of  $\psi$ from specific migration mechanisms, using the Kozai cycles + tidal friction (KCTF) model of \citet{fab07} and the planet-planet scattering model of \citet{nagasawa08} (that also includes the Kozai effect and tidal friction; PSTF) as test models.

We find that conclusions about which migration mechanism is responsible for misalignment depend crucially on the assumption of whether there exists a population of intrinsically aligned systems ($\psi=0$).  Without allowing for an intrinsically aligned population we find that the KCTF model is favored over the PSTF mechanism (\S\ref{analysis}), but allowing for this population we find that PSTF is favored (\S\ref{twomode}), with the most likely fraction of misaligned systems being $\scatfmax$, and between $\scatlowlim$ and $\scathilim$ with $95\%$ confidence.  We also find that this two-mode migration model (PSTF + aligned) is significantly favored over single-mode KCTF migration.  This agrees with \citet{schlaufman10}, who also concluded that there is likely to be an aligned population, based on detecting fewer likely-misaligned systems than expected based on predictions of the KCTF model.  

These results may be an indication of two migration channels for close-in planets, one that acts gently, preserving the alignment  of planet orbits with the stellar spin, and one that acts impulsively, causing misalignment.  The gentle mechanism might well be disk migration \citep{lin96}, and our analysis suggests that the misaligning mechanism is not solely the Kozai effect but rather some mechanism that distributes $\psi$ more broadly, such as planet-planet scattering in combination with the Kozai effect.  This accords with the conclusion of \citet{matsumura10} that some of the close-in planets with non-zero orbital eccentricities are likely to have been formed by planet-planet scattering and subsequent tidal circularization.

Focusing on the subsample of \nhot hot star ($T_{\rm   eff} >6250$ K) $\lambda$ measurements, which \citet{winn10} predict might be the only systems to maintain their primordial misalignments, we find that the data prefer the single-mode PSTF model over KCTF, with a confidence of $85\%$.  We also find that a single migration mechanism is sufficient to describe the current hot star $\lambda$ distribution, without including an intrinsically well aligned fraction.

Looking to the future, we also calculate the number of additional hot star $\lambda$ measurements necessary to achieve $>95\%$ confidence in the hot star model selection (\S\ref{howmany}).  
We find that if either of our single-mode mechanisms describes the $\psi$ distribution around hot stars, then a total data set of about 80-100 $\lambda$ measurements should definitely be sufficient to solidify which is the preferred model, with a $>50\%$ probability of confident model selection with a total data set of only about 40, if scattering is indeed the best explanation for the $\psi$ distribution of close-in planets.



Thus, we suggest that if RM studies wish to be able to identify migration mechanisms through measuring $\lambda$ distributions, they should concentrate on measuring $\lambda$ for planets around hot stars.  Of course it is conceivable that migration mechanisms themselves might be different around different types of stars, in which case hot star $\lambda$ measurements might not tell us anything about cool star migration, but that is the kind of question that will only be able to be explored when much more data is available.

Finally, we emphasize that the results in this paper are illustratory more than definitive, as we have only tested only two very specific misalignment models.  Consequently, we encourage continued theoretical work predicting $\psi$ distributions, as the analysis we present may be applied to any prediction.   We can use the results of this paper as a guide for what to expect from such future analyses.  For example, models that favor larger values of $\lambda$ more strongly than the present KCTF prediction (as does the PSTF model) are likely to be preferred.  We also suggest that an interesting question to pursue would be self-consistent planet formation and dynamical evolution calculations, in order to explore planet-planet scattering in the context of realistic formation scenarios (in contrast to the fixed initial conditions of the \citet{nagasawa08} simulations.  A particularly intriguing angle to explore in such work would be whether a plausible explanation for the observed trend in $\lambda$ with stellar temperature/mass might be explainable by more massive stars tending to form more massive planets in closer proximity, so as to make planet-planet scattering (and thus spin-orbit misalignment) more common among earlier-type stars.  Migration models might also be combined with models that predict that large values of $\psi$ might originate from the host star itself being tilted relative to the disk \citep{lai10}, rather than solely from the effects of planet migration, though current observations suggest that protoplanetary disks tend to be aligned with stellar spins \citep{watson10}. 

Planet migration has been a mystery ever since the first hot Jupiters were discovered, with very little observational evidence to substantiate theories.  As fossil remnants of planet migration, spin-orbit angles are key to understanding the origins of these close-in planets, and the first few years' worth of $\lambda$ measurements are beginning to give substantial clues.  Many more transiting planets will be discovered in the near future thanks to the productivity of transit surveys, and as long as RM measurements of these systems continue, our understanding of planet migration will continue to improve.

\acknowledgments{We acknowledge very helpful suggestions on the structure and content of this paper from Josh Winn and Dan Fabrycky, and useful comments from an anonymous referee.  T.~D.~M.~thanks his office mate Krzysztof Findeisen for helpful suggestions and feedback throughout the course of this project, and also acknowledges the Penn State Astrostatistics Summer School, which helped clarify his thinking about model selection.  J.~A.~J.~thanks Jon Swift, Michael Cushing, Brendan Bowler, Justin Crepp, and Ed Turner for illuminating discussions over the years on topics related to data analysis and statistical methods.  }

\bibliographystyle{apj}
\bibliography{myrefs}

\end{document}